\documentclass[final,5p,11pt,authoryear]{elsarticle}

\usepackage{amssymb}
\usepackage{amsmath}

\usepackage[colorlinks=true, allcolors=blue]{hyperref}
\usepackage{threeparttable,booktabs}

\hypersetup{
  pdfauthor={Mia-Katrin Kvalsund and Mikkel Elle Lepperød},   
  pdftitle={The Generalist Brain Module: Module Repetition in Neural Networks in Light of the Minicolumn Hypothesis},  
}

\journal{Neural Networks}

\begin{document}

\begin{frontmatter}



\title{The Generalist Brain Module: Module Repetition in Neural Networks in Light of the Minicolumn Hypothesis} 


\author[label1]{Mia-Katrin Kvalsund} 
\author[label1,label2]{Mikkel Elle Lepperød\corref{cor1}}
\ead{mikkel@simula.no}
\cortext[cor1]{Corresponding author}

\affiliation[label1]{organization={Department of Physics, University of Oslo},
            state={Oslo},
            country={Norway}}
            
\affiliation[label2]{organization={Department of Numerical Analysis and Scientific Computing, Simula Research Laboratory},
            state={Oslo},
            country={Norway}}

\begin{abstract}
As modern artificial intelligence (AI) reaches new heights in performance and power consumption, there is still no better neural network than the brain. As it stands, biological intelligence remains unmatched in its robustness, adaptability, and efficiency, making the brain a recurring source of architectural inspiration. This review explores one such path: Neural module repetition in AI that is consistent with the minicolumn hypothesis, which sees the neocortex as a distributed system of repeated modules – a structure we connect to collective intelligence (CI). Despite existing literature on facilitating work on the minicolumn, there remains a lack of comprehensive reviews linking the cortical column to the architectures of repeated neural modules. This review aims to fill that gap by synthesizing historical, theoretical, and methodological perspectives on neural module repetition. We distinguish between architectural repetition – reusing structure – and parameter-shared module repetition, where the same functional unit is repeated across a network. The latter more consistently exhibits key CI properties such as robustness, adaptability, and generalization. The reviewed work suggests that the repeated module tends to converge toward a generalist module: simple, flexible problem solvers capable of handling many roles in the ensemble. This generalist tendency may offer solutions to longstanding challenges in modern AI: improved energy efficiency during training through simplicity and scalability, and robust embodied control via generalization. While promising empirical results suggest such systems can generalize to out-of-distribution problems, theoretical results are still lacking. Overall, architectures featuring module repetition remain an emerging and unexplored architectural strategy, with significant untapped potential for both efficiency, robustness, and adaptiveness. We believe that a system that adopts the benefits of CI, while adhering to architectural and functional principles of the minicolumns, could challenge the modern AI problems of scalability, energy consumption, and democratization. With this review, we hope to connect two unfamiliar communities and bodies of literature to facilitate future work toward efficient, green, and general AI. 
\end{abstract}


\end{frontmatter}



\section{Introduction}







Since its inception, Artificial Intelligence (AI) research has been influenced by cognitive science and neuroscience. This symbiotic relationship is evident in models such as the perceptron and the neural network: After all, the only intelligence we know is that of the animal brain and body. It has to be remarkably robust; it has to be able to generalize to novel situations without making mistakes; and above all, it has to be resource efficient. As modern AI reaches new heights in terms of energy expenditure and scale, the fields of biologically- and neuroscience-inspired AI believes that learning from the brain only becomes more and more important \citep{zador2022toward}. Therefore, inspired by a recent synthesis on collective intelligence \citep{mcmillen2024collective}, this review goes somewhat outside of main-stream AI to explore the topic of brain-inspired collective intelligence, grounded in a real biological neural architecture.

\begin{figure*}
    \centering
    \includegraphics[width=\linewidth]{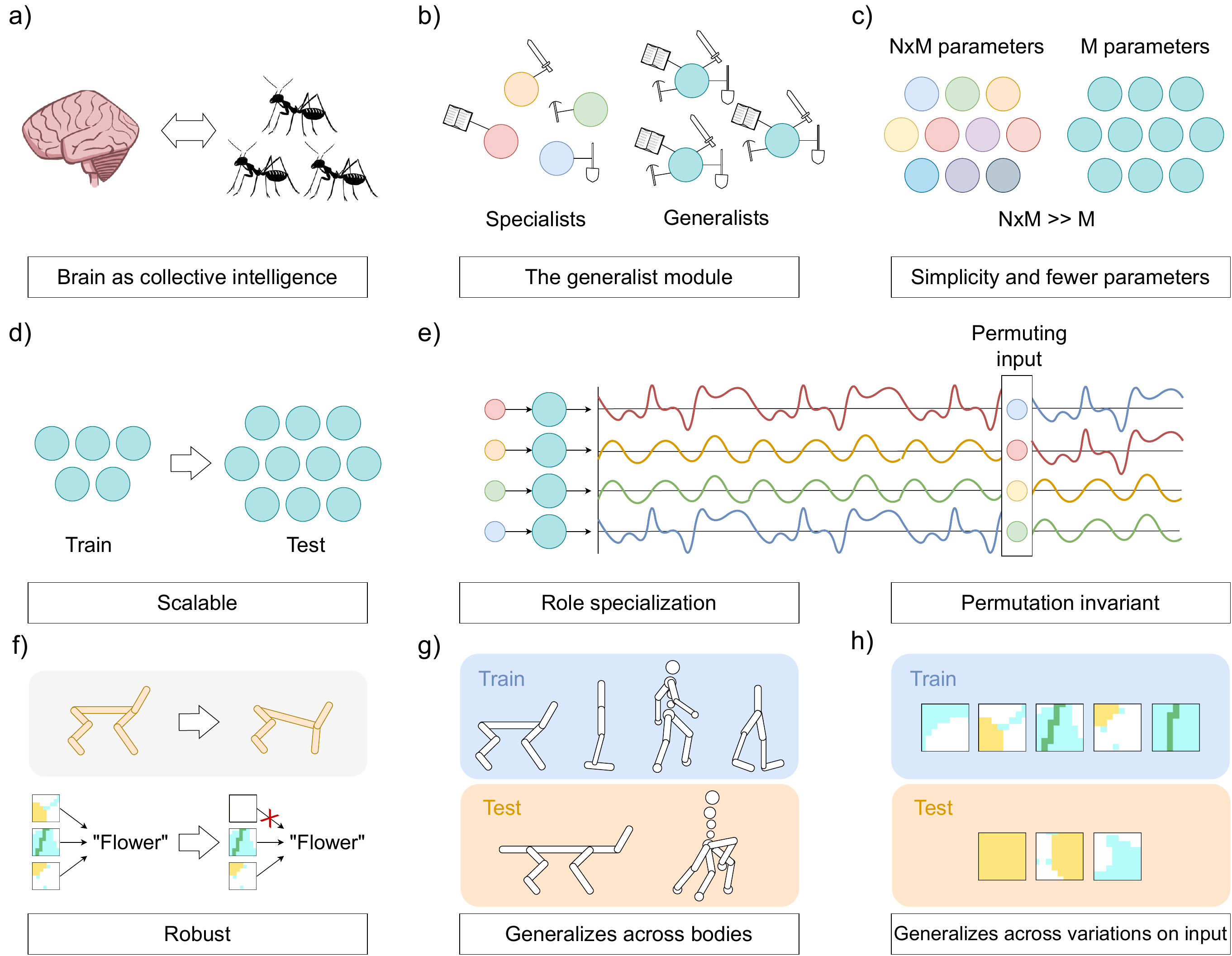}
    \caption[Key theoretical and empirical insights on the generalist module]{\textbf{Key theoretical and empirical insights on the generalist module:} a) We conceptualize the neocortex as a system embodying collective intelligence through its distributed, homogeneous column architecture. This suggests that systems like it can be analyzed using the literature on swarm and collective intelligence. b) Each module in the collective is a generalist (right), one that can flexibly adopt different skillsets as needed by the collective, unlike systems with fixed-function modules (left). c) In a system of N modules and M parameters per module, generalist architectures reduce total parameters to a lower bound of M, by sharing a single parameter set. d) Because the modules are defined with a common parameter set and trained to work with copies of itself, it can potentially scale up after training. This could allow training on a smaller version of the model, and scale up after deployment. It can also better allow multi-task generalization, where number of inputs are different between tasks. e) The generalist module may specialize to only perform one function within an episode, given its specific input (left). Various works also show that the modules can switch roles on permuted input (right), given that the downstream architecture is similarly permutation invariant. f) Top: As a function of learning to work in many positions in the body of a robot, the generalist module may recover function when limbs are removed. The stump module may change their role in response to their new inputs. Bottom: Similarly, a classifying system may still maintain function as some visual fields are removed. g) Through the same logic, the generalist module has been shown to generalize across out-of-distribution bodies when trained on multiple bodies during training. h) This is a general pattern of being able to generalize across different inputs it may get in any given task, which again makes it more scalable. Figure e is adapted from the findings in \protect\cite{pedersen2022minimal}; figure g is adapted from the findings in \protect\cite{huang2020one} (the bottom two robots are from their supplementary videos and the top robots are adapted from MuJoCo \protect\citep{todorov2012mujoco}).}
    \label{fig:concept_figure}
\end{figure*}

The minicolumn hypothesis debuted in the 1970s with the work of Mountcastle in his extensive corpus on the neocortex (\cite{mountcastle1978organizing, mountcastle1997columnar}). It was introduced as an elementary brain module, repeated all throughout the neocortex to make up most higher-order thinking and sensor-processing. It promised that by understanding it, the whole brain could soon be understood by the sum of its parts. However, after many years, the usefulness of working with the column would be criticized, as some authors meant that the column as an anatomical entity served no specific purpose \citep{horton2005cortical}. It was therefore exciting when in \citeyear{hawkins2019framework}, \citeauthor{hawkins2019framework} published their Thousand Brains Theory, which would be named as such in the 2021 book called "A Thousand Brains" \citep{hawkins2021thousand}. This theory focused on ascribing a function to the cortical column, a framework to continue work on the column hypothesis. This followed the recent findings that grid cells were found in the neocortex \citep{long2021novel}, and that columns, therefore, may have a sense of space. \citeauthor{hawkins2021thousand} defines the cortical column as one of the "thousands of brains" repeated throughout the neocortex, each capable of creating a model of complete objects. The columns communicate locally with other columns and with areas involved in motor behavior. This parallel processing allows for rapidly reaching a consensus on objects, to form beliefs about the world in a continual loop of sensory-motor learning. 

In practice, a system inspired by the minicolumn-hypothesis should have at least one neural module that is repeated in a closed system, akin to how the cortical column is repeated throughout the neocortex. The repeated module can have parameter sharing or differentiated parameters. Furthermore, the ensemble must be connected, either through message-passing, output integration, or through a mechanical body. This excludes swarm and multi-agent robotics, because it concerns modules in separate mechanical bodies. Our focus is on modular alternatives to conventional, non-modular neural networks, where computation is entangled in a single structure. For brevity, we refer to these as \textit{monolithic} networks.   

The minicolumn hypothesis is an exciting hypothesis for biologically inspired AI. As Mountcastle himself did \citep{mountcastle1978organizing}, we can see the columns as a distributed system of repeated neural modules. As we will show in section \ref{seq:swarm}, this qualifies the columns to be interpreted as collective intelligence. By drawing this parallel, we can hypothesize about the potential benefits of AI inspired by the minicolumn hypothesis, drawing on the literature from distributed systems and Swarm Intelligence (SI). 

SI and distributed systems have many useful properties stemming from their multi-scale competency -- meaning that they display competencies on the ensemble level and every level below (such as the agent level, cell level, etc.) \citep{mcmillen2024collective}. These properties have been summarized in Figure \ref{fig:concept_figure}. Principally, the swarm agent is a simple and flexible computational unit that works to achieve the ensemble's goal \citep{Dorigo2007}. Because of its simplicity, it requires far fewer parameters than the ensemble as a whole, which can be resource-effective. Further, because it can subsume any role the ensemble needs, it displays a sense of general problem-solving, making the ensemble adaptable and generalizing, even at an economic scale. These properties are not only more biologically plausible than monolithic networks; they could also challenge the longstanding problems of AI, such as its energy consumption during training \citep{bashir2024climate, strubell2019energy, schwartz2020green} and scalability \citep{thompson2020computational, bender2021dangers}. Moreover, as a potentially more cost-effective alternative, it helps with democratization of AI \citep{schwartz2020green}. 

\begin{figure}
    \centering
    \includegraphics[width=\linewidth]{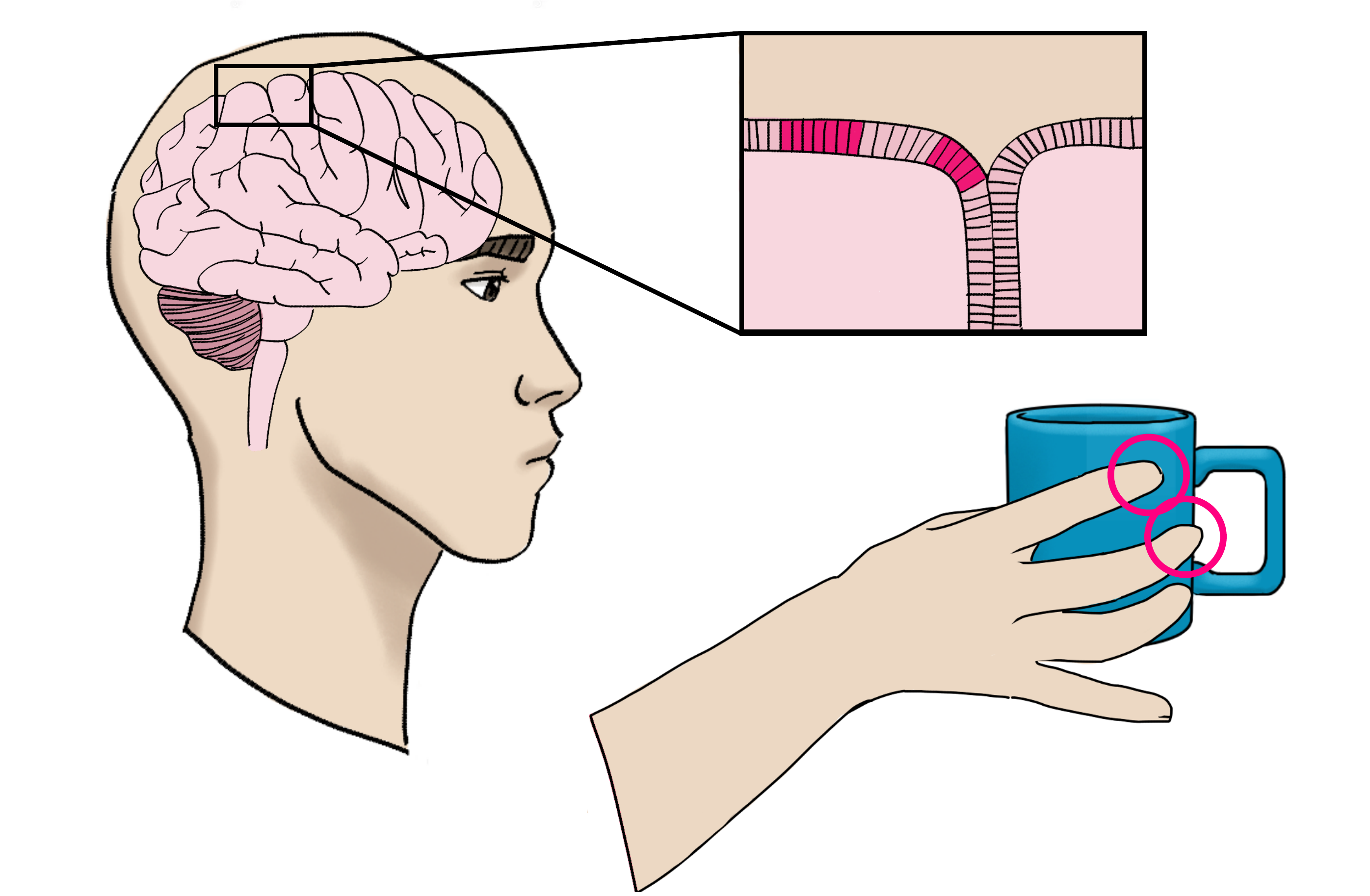}
    \caption{\textbf{Cortical columns in the sensory cortex each receive a limited part of the senses.} Across the neocortex, the outer part of the mammalian brain, there are small units of neurons called cortical columns. They correspond to only a small part of the body, such that the whole sensory cortex corresponds to the whole body.}
    \label{fig:neocortex}
\end{figure}

Furthermore, the column is closely connected to sensor-processing and motor control, and therefore offers an exciting direction for robot control that bears the benefits associated with SI. In the sensory cortex, columns will map to a limited patch of the body, receiving sensory information and contributing to the control of this patch (see figure \ref{fig:neocortex}). Therefore, this framework is somewhat unique among distributed control, as it proposes a way to have distributed control of a heterogeneous and non-modular body through integrating column outputs. As the race for humanoid robots continues, and AI researchers propose more embodied tests of intelligence \citep{zador2022toward}, this distributed and biologically inspired approach could hold some promise. 

There is a lot to be gained from considering the analogue between the minicolumn hypothesis and collective intelligence. Yet, previous work inspired by cortical columns has not been aware of its collective intelligence roots, herein its advantages and disadvantages. Likewise, it is uncommon for AI research featuring collective intelligence to be aware of its neural analogs. For example, the work of Hawkins and his lab, Numenta, excellently models columns through the Hierarchical Temporal Memory model (HTM) and their new model, Monty, from the Thousand Brains Project \citep{george2009towards, clay2024thousand}. These systems can be classified as collective intelligence - yet, throughout their literature, collective intelligence theory is not used to enrich their theories or models. Likewise, works on embodied robots with column-like control \citep{pathak2019learning, huang2020one, ferigo2025totipotent}, reviewed here, is not concerned about how the architecture of the neocortex can inform their own architectural choices, and choice of tasks. More importantly, these separate sub-fields are largely not aware of each other's work: Perhaps with some forced cross-pollination, progress on both ends might accelerate.

Previous reviews have, to the best of our knowledge, not discussed neural module repetition in the same context or depth as this review. While \cite{amer2019review} discusses "modular node topology", which corresponds to module repetition,  in the context of modular neural networks, this is not the focus of their review. Similarly, the excellent review of \cite{ha2022collective} discusses works that display collective intelligence, herein many of the works discussed in this review, but has a more historical perspective on deep learning and collective intelligence as opposed to our focus of providing a synthesis of the method across fields. Numenta's own review to facilitate work on the "thousand brains" theory of intelligence goes through the theoretical context of why such a direction is needed now, namely that we are still not much closer to general AI \citep{hole2021thousand}. Their review is understandably focused on Numenta's work, and suggests that their own model, the HTM, could be a way forward.     

\begin{figure}
    \centering
    \includegraphics[width=0.78\linewidth]{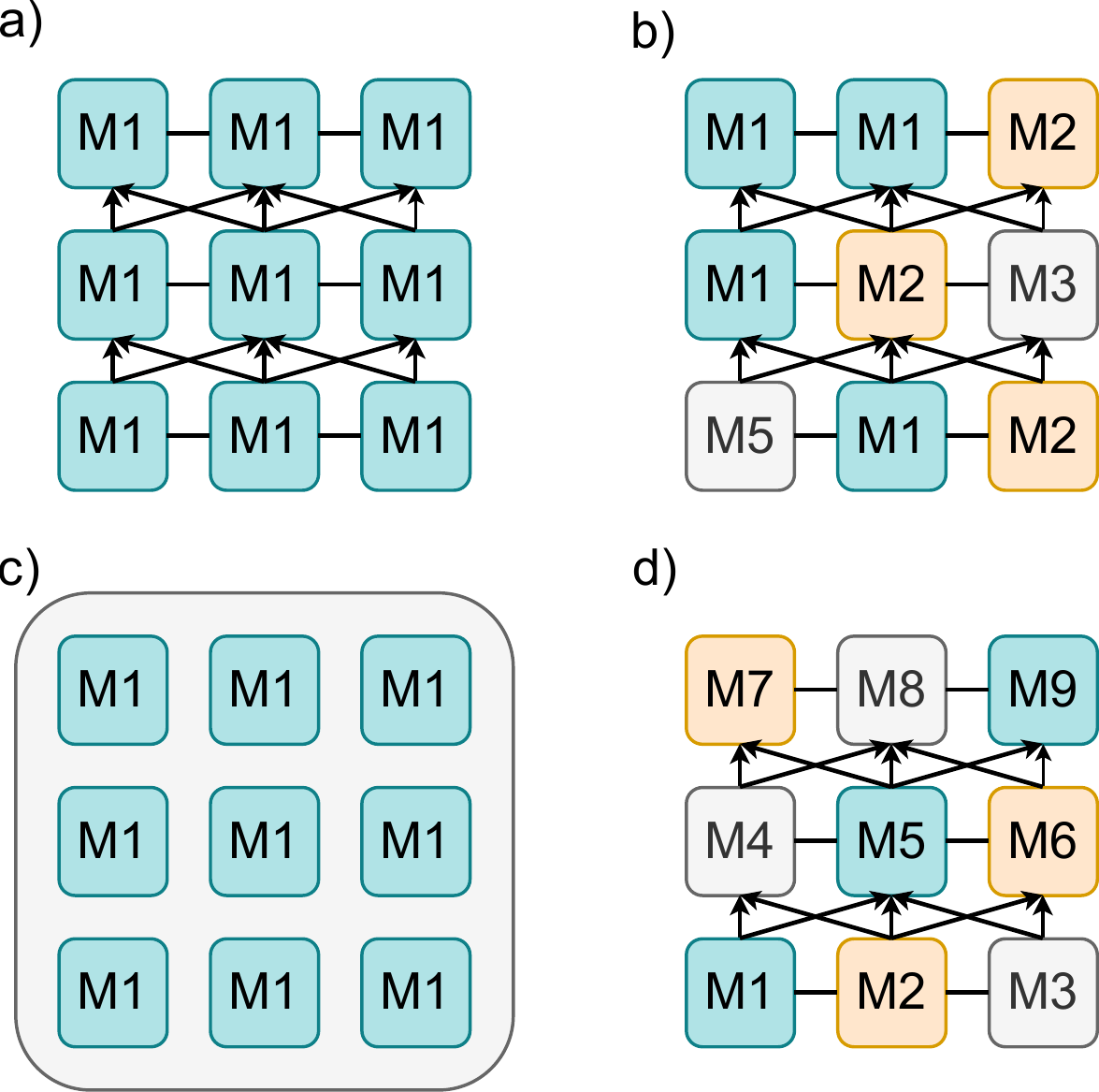}
    \caption{\textbf{Examples of module repetition this review will concern.} The boxes labeled M are modules, where colors signify that a specific architecture is used. \textbf{a)} Network of parameter sharing modules: A network repeating one module, M1, in multiple positions. \textbf{b)} We also include when repetition of modules is not fully homogeneous. Here, a network repeats two modules, M1 and M2. \textbf{c)} Disconnected and parallel modules with parameter sharing connected by some system. The system can be message passing, or part of a mechanical body. \textbf{d)} Architectural module repetition: All the module parameters are different, but the architectures are repeated, indicated by the matching color.}
    \label{fig:examples}
\end{figure}

This review seeks to facilitate AI research inspired by the minicolumn hypothesis. Specifically, we will review ensembles of neural network modules where at least one defined module is repeated in a closed system (Figure \ref{fig:examples}). However, as opposed to previous reviews, we seek to not only discuss the methods used and their benefits, but also present them along with the relevant concepts to understand their potential fully -- especially the theoretical benefits of such an architecture. Further, we catalog to what degree such benefits have been observed in the literature. The reviewed work is a representative sample such that as many methods as possible is showcased, but we will take special care to include research on robot control because this will give us insight into how these methods affect the body. By presenting a focused synthesis, we hope that the following discussion will help and inspire AI researchers seeking to make systems inspired by the minicolumn hypothesis themselves. Secondly, we hope to inspire collective intelligence research to explore new avenues of SI capabilities. 

In section \ref{seq:cognitive_science}, we review how the concept of a module in AI inherited different meaning from the fields of cognitive science and neuroscience. Section \ref{seq:theoretical} relays theoretical advantages and disadvantages of neural module repetition. We then review the disparate works that employ neural module repetition in section \ref{seq:methods}, before we summarize the benefits as observed in the included work in section \ref{seq:observed}. Finally, section \ref{seq:conclusion} summarizes the history and concludes. 

\section{Cognitive science, AI, and the modularity of mind}
\label{seq:cognitive_science}

In the fields of AI research and engineering, the view of a neural module is not well defined. Some consider modules as being domain-specific and having information encapsulation; others implicitly attribute regularity to modularity \citep{lipson2007principles}. Most articles seem to consider modularity without regularity, where the module is a specialized and sparsely connected unit. In fact, the comprehensive review on modular neural networks (MNNs) by \citeauthor{amer2019review} (\citeyear{amer2019review}) only considered sparsely connected modular networks with specialized functions, indeed because this reflects the state of MNN research. Still, other fields or parts of the literature focus more on homogeneous modules, as we will later show. The difference between these two focuses is between the idea of an expert module and the generalist module. This split in the literature between specialized experts and repeated generalists has a counterpart, and likely influence, in cognitive science. For context, a brief overview will be given: Keep in mind this is a debate rich in literature and opposing opinions (interested readers might read \cite{eisenreich_control_2017, zerilli_neural_2019, egeland2024making, mccaffrey2023evolving}). Due to the interdisciplinary nature of the following discussion, we will use the term "module" as it makes sense in an AI context: A building block of (artificial) neurons with sparse connections to surrounding neurons. Similarly, the term "distributed" will in this review be used to mean decentralized, lacking centralization/functional localization, and does not oppose modularity.

\begin{figure*}
    \centering
    \includegraphics[width=\linewidth]{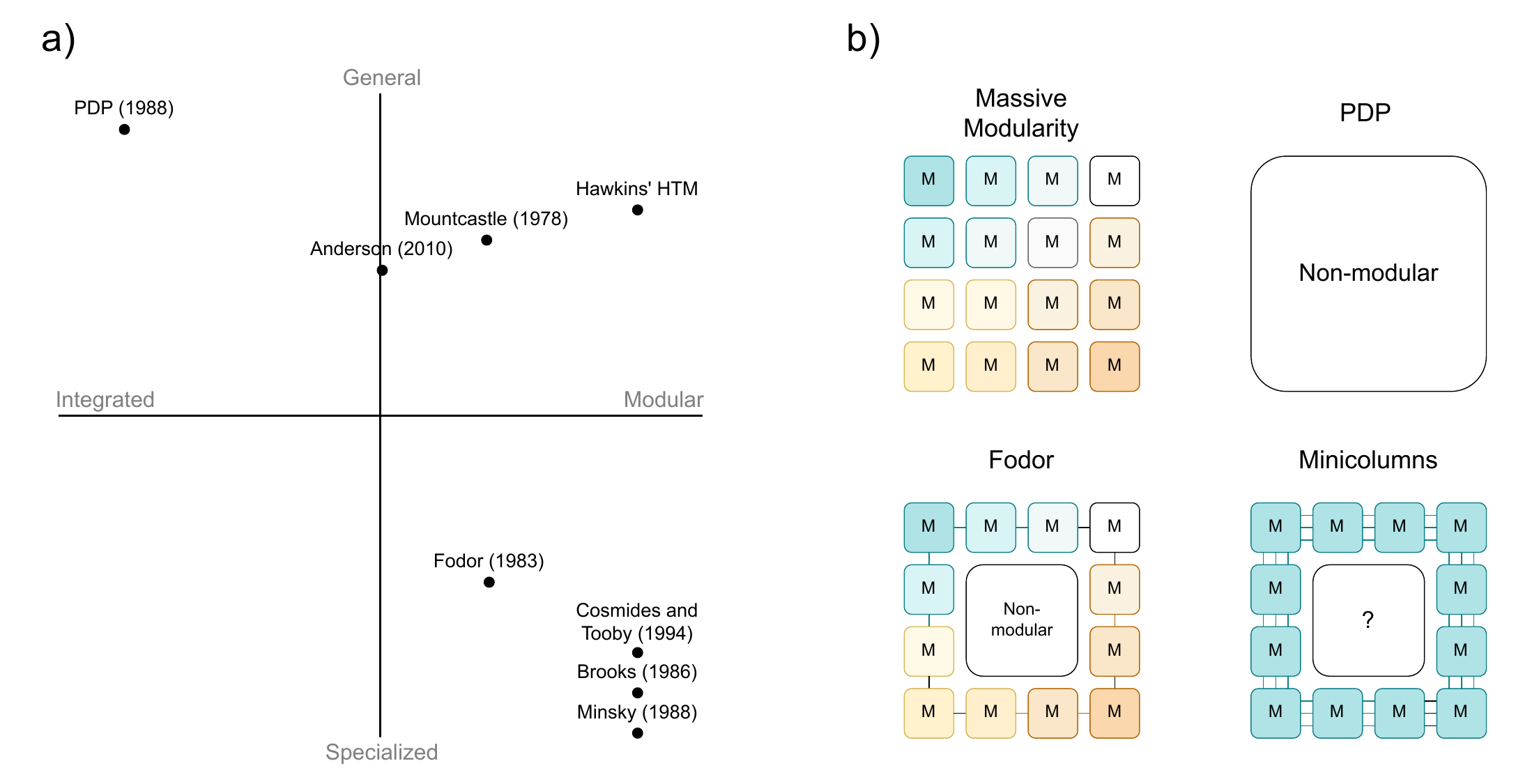}
    \caption{\textbf{Views on the brain's modularity.} a) The various authors described in this section fall somewhere on the spectrum between believing in specialized versus general and modular versus integrated cognition. b) Each box marked M represents a brain module, with the color signifying functional specialization. As such, both Massive Modularity and Fodor saw a functionally diverse brain. While the same is true for the minicolumn hypothesis, is assumes that the modules share some common function, and does not make assumptions about the rest of the brain. PDP straightforwardly thinks of the brain as non-modular.}
    \label{fig:cognitive_science}
\end{figure*}

The beginning of modern brain science saw an emerging belief that the brain was both anatomically and functionally modular (as exemplified by neuroscience works such as Broca's area (1860s) and Wernicke's area (1870s) that showed localized function). Perhaps the most famous case of this is that of phrenology, otherwise known as organology, where regions of the cranium were used to measure faculties of the human psyche with measuring tape \citep{hytche1846anatomy, greenblatt1995phrenology}. Though rejected by the middle of the 1800s \citep{greenblatt1995phrenology, jones2018empirical}, the idea of functional modules continued to be appealing. 

In the 1980s, a confluence of several factors came to a head, and the debate of the brain's modularity was seriously reopened with the book \textit{Modularity of Mind} by \citeauthor{fodor1983modularity} (\citeyear{fodor1983modularity}). He posited that the brain was composed of specialized modules with innate functions formed through evolution. However, Fodor restricted modularity to the newer brain regions. Furthermore, Fodorian modules are both localized and encapsulated \citep[p. 36-37]{fodor1983modularity}, which later authors would move away from \citep{cosmides1994origins}. Authors such as \citeauthor{cosmides1994origins} (\citeyear{cosmides1994origins}) punctuated that because specialized modules were modifiable by evolution without disturbing other functions, they would necessarily be favored in evolutionary selection throughout the whole brain. Consequently, they established the massive modularity hypothesis, the notion that the entire brain was modular. The views of massive modularity proponents and Fodor were in internal opposition, where the degree of innateness, information encapsulation, and distributedness were being negotiated \citep{fodor2000mind, pinker2005so, frankenhuis2007evolutionary}. Still, Fodor and Massive Modularity exemplify the view that dominated: the brain consists of specialized modules, their functions are innate, and are singularly modifiable by evolution. 

The AI community was following a similar thread of specialized modules around these times. The prominent AI researcher Marvin Minsky released the book \textit{Society of Mind} in 1986 \citep{minsky1988society}, taking a modular view of the brain that was somewhat of a step further than Fodor's \textit{Modularity of Mind}. In \textit{Society of Mind}, Minsky treats all brain functionality as autonomous, communicating expert "agents". Two years before, the prominent roboticist Rodney Brooks released his "subsumption architecture" \citep{brooks1986robust}, where he also argued for distinct agents performing sub-tasks, though in a more hierarchical way. Brooks would later tie his architecture to Minsky's \textit{Society of Mind} \citep{brooks2014build}. 

Connectionism and the PDP (parallel distributed processing) movement opposes Massive Modularity and Fodor. Here, the brain is viewed as a distributed system of neurons that self-organize to perform some higher brain function. See the review by \citeauthor{eisenreich_control_2017} (\citeyear{eisenreich_control_2017}) for arguments on how a distributed brain could accomplish rather centralized functions. Proponents of these theories are the connectionists or the PDP movement, a group that was, since its inception, always concerned with their theories' application to AI and cognitive science. The influential PDP book by Rumelhart, McClelland, and their PDP research group \citep{rumelhart1988parallel} notably also appeared in the 80s, a few years after "Modularity of Mind". 

Much of the debate of the 80s happened before it was truly clear how plastic the brain could be. That is why, when brain regions were shown to be reused across functions in what is called neural reuse, cognitive scientists began to consider the effect of neural reuse on existing theories of brain architecture – and found that domain-specificity could not be maintained \citep{anderson2010neural, anderson2007massive}. For example, The Massive Redeployment hypothesis, fronted by \citeauthor{anderson2007massive} (\citeyear{anderson2007massive}), states that brain regions are reused in many brain functions, and are thus more generalist than previously thought. Importantly, the distinction comes from the fact that a specialized module supports only a few functions and therefore evolves specialized wiring and supports information encapsulation; while a generalist module has to support many functions and therefore withers information encapsulation as it becomes more general. While neural reuse does not necessarily oppose the modularity debate, authors specifically deny the existence of functionally specialized/domain-specific modules and information encapsulation \citep{anderson2010neural}.

While neural reuse complicates the idea of specialized modules, there is still one module that might survive this evidence: Mountcastle's cortical column. A few years before the modularity debate was restarted, \citeauthor{mountcastle1978organizing} (\citeyear{mountcastle1978organizing}) wrote \textit{"An organizing principle for cerebral function: the unit module and the distributed system"}. Coming in before the Fodorian module was defined, he presented an anatomical mini-module that was a generalist, repeated module: He described a neocortex that consists of metamodal and anatomically selfsame columns of neurons that functioned in a distributed system \citep{mountcastle1997columnar} (though how anatomically similar they are is often debated \citep{amorim2010whose}). This hinges on ontological plasticity - that virtually identical brain modules can, according to their context (inputs/sensors), change their function to accommodate new wiring (new sensors) \citep{zerilli_neural_2019}. Thus, the hypothesis entailed architectural generalist modules that could become life-time specialists through plasticity. Although these ideas were formulated early on, this notion of a cortical mini-module did not have a large impact on AI before recently. In the years leading up to 2021, when Jeff Hawkins debuted his "Thousand Brains Theory" \citep{hawkins2019framework, hawkins2021thousand}, columns were just starting to see a revival (largely due to his Hierarchical Temporal Memory model (HTM) \citep{george2009towards}). Moreover, cognitive science researchers such as \citeauthor{zerilli_neural_2019} (\citeyear{zerilli_neural_2019}) would start to adopt the column as perhaps the only contender for "brain modules" that survived the evidence of neural reuse. 

In summary, cognitive science saw the reintroduction of modularity in the 80s and grappled with the extent to which the brain was modular and what the modules were. At the same time, AI researchers like Minsky and Brooks formed influential theories of how to shape artificial minds, and the PDP movement began with its view of the distributed brain. All the while, work on Mountcastle's cortical columns piled up. Through a series of preceding papers, the Thousand Brains Theory was published in 2021, which argued for a modular yet distributed approach featuring neural module reuse at a large scale. In short, cognitive science has not come to any conclusions regarding modularity despite the strong stances available, and the AI community has inherited a general confusion as to what neural modules are, and what they can do.

While AI research at large seems to align with either specialized modules (like Minsky and Brooks, and more recently also Friston \citep{friston2024designing}) or PDP principles (deep learning is continuing the connectionist philosophy \citep{berkeley2019curious}), this review concerns works that employ generalist modules in neural networks. To that end, we include works featuring both modularity and regularity, though this necessitates including repeating module architectures without parameter sharing. Throughout the AI literature, the phenomenon of a truly regular modular network is somewhat of an oddity. It appears either in connection with Mountcastle's columns or to collective intelligence and usually in connection with Evolutionary Computation.

\section{Theoretical advantages and disadvantages of repeated module architectures}
\label{seq:theoretical}

\begin{figure*}
    \centering
    \includegraphics[width=0.7\linewidth]{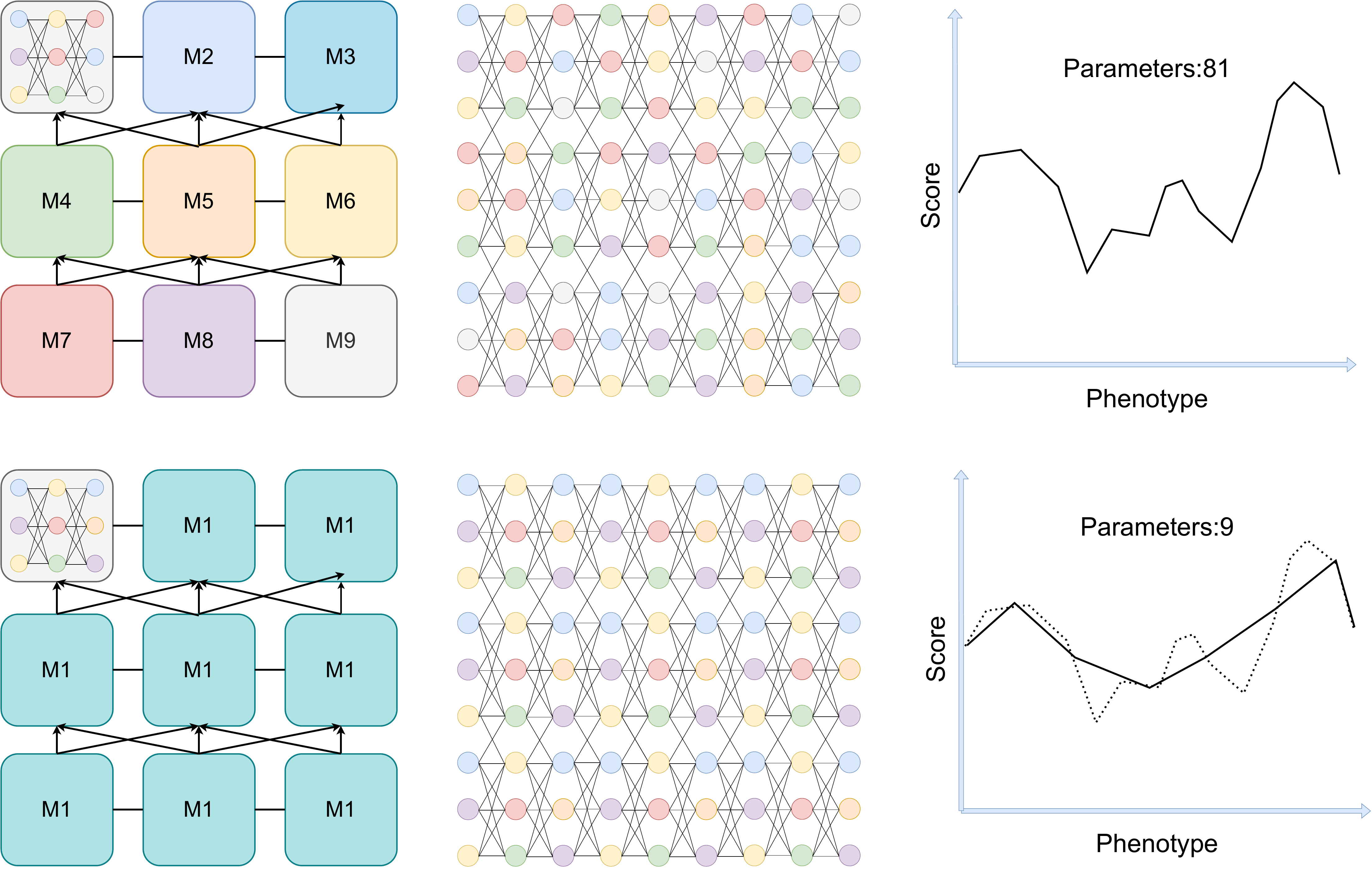}
    \caption{\textbf{An example of a neural network with 81 weights without neural module repetition (Top) and with (Bottom).} The graphs are the imagined search spaces, with the possible networks (Phenotypes) on the bottom axis and the score of the possible networks on the vertical axis. In the normal case, 81 parameters means an 81-dimensional search space (here collapsed on one axis labeled Phenotype) of real-valued axes. In the bottom case, with module repetition, the search space is only 9-dimensional. The resulting coarser sampling as dimensions are reduced is depicted as the solid line sampling only some parts of the dotted line (the 81-dimensional case).}
    \label{fig:search_space}
\end{figure*}

Before delving into projects from AI, this section outlines the theoretical advantages and disadvantages of neural module repetition in neural networks. Advantages and disadvantages have been collected from literature on cognitive science, AI, neuroscience, swarm intelligence, and modular robotics. Then, after reviewing articles in the next section, we will revisit the advantages and disadvantages as they have been observed by the AI community.

\subsection{Fewer, repeated parameters}

No matter how regular and modular neural repetition is employed, a reduction in parameters will follow. If $N$ is the number of modules and $M$ is the number of times they are repeated, the general rule is $N \times M$ parameters for a non-repeating network and $N$ for a repeating network. For example, for systems with homogeneous repetition, where the module has $N=9$ parameters and is repeated $M=9$ times, the repeat case will have a minimum of 9 parameters and the non-repeat case will have a minimum of 81 parameters (Figure \ref{fig:search_space}). Reducing parameters means a reduced search space and a smaller job for the optimizer, which can result in shorter optimization, saving time, money, and energy.

Moreover, for optimization methods that handle large search spaces poorly, like Evolutionary Computation, reducing parameters is one way to get better results without sacrificing network size. When reducing the search space with a useful heuristic, Evolutionary Computation in robotics is often shown to perform better than in the full parameter search space \citep{Veenstra2020, hornby2003generative}. The heuristic will expand on sparse parameters by some rule to create a network. In our case, the rule is repetition in a predetermined or optimized blueprint. 

However, reducing parameters takes away fine-grained control over the final phenotype. For example, the discussion on domain-specialized versus domain-general systems raises a valid argument against the domain-general module, called the "debugging" problem \citep{zerilli2017against, cosmides1994origins}. This problem can be summarized as such: If two functional modules 1 and 2 share resources/parameters $P$, how can one function be optimized without changing the other? Changing $P$ with regards to function 1 seems to inevitably degrade function 2. As such, we lose the ability to fine-tune for either function, and must settle for a lesser solution that serves both functions. However, the brain does not suffer greatly from its evident neural reuse \citep{anderson2010neural, anderson2014allocating}. This gives us an indication that though the lack of fine-control logically is an issue, systems can still excel with this constraint. Consequently, the "debugging" issue should be kept in mind when constructing such systems. Moreover, as we will see in the following section, this constraint comes with a possible strength. 

\subsection{Generalization is tricky}

Imagine that you are about to be hit by a tram. This is a novel situation to you, yet you know exactly what to do: You leap out of the way and you escape with your life. How did you manage to solve a situation you've never encountered before? 

This question, based on the work of \citeauthor{cosmides1994origins} (\citeyear{cosmides1994origins}), is one of generalization: How can life-time learning, or optimization, act on behavior that may only be seen once during a creature's lifetime? In a domain-general system, such as in traditional deep learning, the problem takes on a holistic representation and often struggle to be broken down to pre-learned behavior. The scenario must be trained for, making the system data hungry, and prone to undergeneralizing. Meanwhile, a modular system of specialists can break the issue down to a few domain-specific behaviors, such as "threat detection", "object avoidance" and "awareness". These capabilities, honed through evolution and practice in less severe circumstances, can rapidly be \textit{composed} at that crucial moment to solve novel challenges, such as "tram avoidance". 

\begin{figure*}
    \centering
    \includegraphics[width=\linewidth]{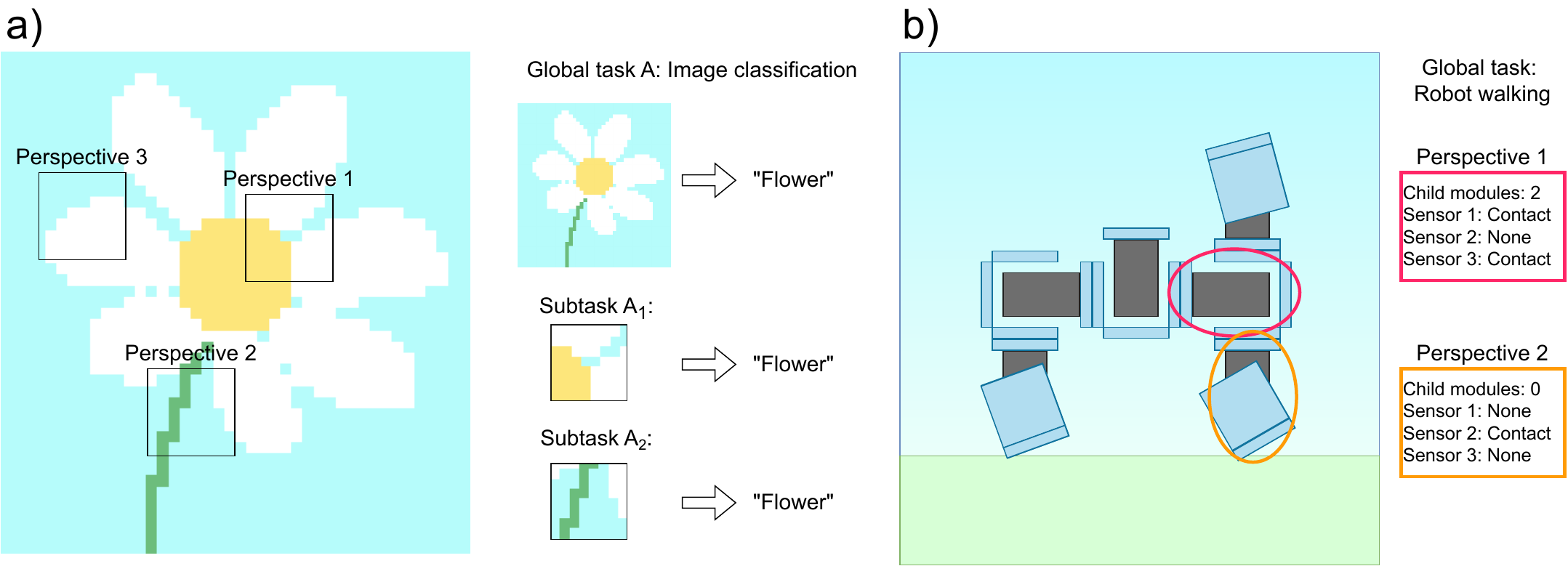}
    \caption{\textbf{Examples of perspectives on tasks.} a) In image classification, pixel neighborhoods can be seen as perspectives on the task. Each perspective-label pair represents a subtask. The parameter set $P$ is encouraged to output such that the global behavior classifies the image. b) In modular robot walking, each physical module has a unique perspective on the task, such as sensor input and connection details. Each module will contribute with behavior that creates the global behavior of walking forwards. Robot adapted from \citep{kvalsund2022centralized}.}
    \label{fig:perspectives}
\end{figure*}

This is the problem of compositionality, which remains a hard challenge for connectionist systems \citep{ruis2020benchmark, goyal2022inductive}. Evidently, a system of repeated generalist \textit{modules} could also fail to compose, for the same reasons, given that they don't specialize within their life-time. However, avoiding a tram can also be seen as a slight variation on avoiding an oncoming bike - a distribution shift. Being robust to such slight differences in situations can also provide generalization. In such situations, repeated modules might have an edge. 

Learning under distribution shifts can encourage generalization, through learning an approximate causal model of the data generating process \citep{richens2024robust}. This is provably true for a monolithic system training on several tasks $A$ to $Z$. We can extend this result to a testable hypothesis for repeated generalist systems: Given that each module shares one set of parameters, receives one coherent subspace as input, and contributes to global action, we can expect a module to learn a causal model of the data generating process. The coherent subspace amounts to a \textit{perspective} on task $A$, such that, for each perspective 1 to $N$, each module must solve a subtask $A_1$ to $A_N$ (Figure \ref{fig:perspectives}). The perspectives thus amount to distribution shifts, which suggests that the modules can generalize to more perspectives on task $A$. This is similar to how metamodal learning has been shown to be beneficial for generalization \citep{lu2023theory}, although the generalist system might have several perspectives within each modality as well.  

This notion of perspectives is similar to how cortical columns in the sensory cortex receive subspaces of a sense as input (Figure \ref{fig:neocortex}). Through this "pinhole" view of the world, \cite{hawkins2021thousand} theorize that the column will still learn complete models of objects. Although Hawkins' and Numenta's conception of "models of objects" does not wholly align with the formal idea of a "causal model of the data generating process" \citep{richens2024robust}, there are important overlaps. Primarily, as an object is sensed through thousand of "pinholes", the columns will get time series of sense data. To recognize this time series through many perspectives at once, in many different orientations and positions, the columns may very well have to infer the underlying causal structure – the object itself – that generates the sensory data. In this sense, learning an object model could implicitly require learning an approximate causal model. Empirically, the reviewed works often showcase generalization, suggesting that there could be truth to the above hypothesis. Furthermore, even though it is unclear whether such generalization could extend to other tasks than task $A$, we observe in the reviewed literature that the generalization can extend to very novel versions of task $A$. 

\subsection{Architectural constraints}

In lieu of variable inputs and outputs, a repeated module has a defined interface. This naturally constrains the system: given a module with input $n$ and output $c$, it is not necessarily trivial how to connect modules to achieve system inputs $N$ and outputs $C$. For example, Convolutional Neural Networks are often constructed with several pooling layers and non-modular, fully connected layers to reduce dimensions and aggregate information appropriately \citep{goodfellow2016deep}. Additionally, the neocortex columns also project to subcortical structures, like the thalamus, in what some believe to be an integrative function \citep{suzuki2023deep}. In general, we observe that the \textit{integration} of the final module output can favorably be non-modular or a pooling operation (see f.ex. \cite{gholamalinezhad2020pooling} for a thorough list on pooling operations beyond max and average pooling). 

While initially constraining, modules also grant a powerful flexibility to number of system inputs and outputs. Monolithic networks such as feed forward neural networks (FFNNs) are constrained to their initial dimensions, and new inputs and outputs typically cannot be added without retraining. The parameter set will correspond to $N$ input neurons, meaning that if one more input is added, at least $H$ new parameters must be trained, where $H$ is the number of hidden nodes. This means that FFNNs cannot have a flexible number of inputs and outputs. For repeated module architectures with parameter sharing, however, modules can be added as needed without retraining: The parameter set corresponds to one module, repeated over the input $N$. If one more input is added, at least one more module can be added, which does not increase the parameter set. If this works zero-shot is up to the integration - we will see empirical results in section \ref{seq:observed} of this working well. Integration can then easily be handled by pooling or adaptive pooling. This can be especially useful when the system is trained on multiple tasks, such as when attempting to achieve multi-task generalization (as discussed in \cite{pedersen2024structurally}).

When using parameter sharing, repeated module architectures can cause issues when it comes to differentiating functions between modules. \cite{pedersen2024structurally} coined the term \textit{Symmetry Dilemma}: Symmetric networks, i.e. networks with parameter repetition, will have several useful benefits. However, if such symmetry is taken too far, the network will become unable to propagate information. In other words, if deterministic modules receive the same input, they will generate the same output, making repetition detrimental to its function. The key lies in finding a balance between repetition and differentiation. Therefore, modules must always be somehow differentiated, for example, by receiving different inputs or by stochastic updates. Often, this can be solved by having each module receive a subspace of the input, such as with perspectives (figure \ref{fig:perspectives}). 

\subsection{The swarm-like properties of a distributed system}
\label{seq:swarm}

Given the description of Hawkins et al. of thousands of repeated neural modules working in parallel in a distributed system, it is apt to compare the idea to Swarm Intelligence. Swarm Intelligence is a field that largely concerns the study of emergent intelligence from swarm systems, usually insects and popularly ants and bees, but social animals also fall under this category \citep{chakraborty2017swarm}. In terms of artificial systems, the study of swarms is abstracted into principles that concern swarm systems in general, including biological and natural systems, as well as artificial \citep{bonabeau1999swarm}.

\citeauthor{Dorigo2007} (\citeyear{Dorigo2007}) defines a swarm system, agnostic to the type of creature involved, as such:
\begin{quote}
    \begin{itemize}
        \item "it is composed of many individuals;
        \item the individuals are relatively homogeneous (i.e., they are either all identical or they belong to a few typologies);
        \item the interactions among the individuals are based on simple behavioral rules that exploit only local information that the individuals exchange directly or via the environment (stigmergy);
        \item the overall behavior of the system results from the interactions of individuals with each other and with their environment, that is, the group behavior self-organizes."
    \end{itemize}
\end{quote}

Here, "individuals" can be thought of as Minsky's agents, single neurons, or the cortical column. In the case of the cortical column, Mountcastle likened the neocortex to a distributed system, especially in the homotypical areas \citep{mountcastle1997columnar}, which we can extend to mean a swarm system. We observe that the neocortex is composed of many columns, they are rather homogeneous, they interact locally, and the overall behavior emerges from their interaction. \citeauthor{mountcastle1997columnar} (\citeyear{mountcastle1997columnar}) adds that they are rather robust to damage because lesions in one area might not fully remove a function. Given that the system of repeated neural modules fits the description of a swarm, we can expect to see swarm properties in the literature to be reviewed. These properties are:

\begin{itemize}
    \item Simplicity: An individual is simple and cannot explain the properties of the swarm on its own \citep[p. 6]{bonabeau1999swarm}.
    \item Flexibility: Division of roles are plastic; individuals can change their role if the swarm requires it. An individual will still specialize for a role it frequents \citep[p. 110]{bonabeau1999swarm}. 
    \item Robustness: Swarms can easily bounce back from disturbances (such as removal of individuals, removal of an entire task-specific group, new environment, etc.) because of their flexibility. Individual elasticity causes swarm resilience \citep[p. 111]{bonabeau1999swarm}. Self-organization causes the swarm to autonomously self-repair its function \citep{Dorigo2007}.
    \item Scalability: Following from robustness, the swarm can also easily scale up \citep{Dorigo2007}.
    \item Parallel execution: The swarm performs many actions in parallel, creating an emergent behavior \citep{Dorigo2007}.
\end{itemize}

Some of these properties follow logically from the benefits outlined above. For example, generalization will allow modules to take on different roles. Other properties come from the architecture: A distributed system, trained in an ensemble, tends to develop an apathy towards the size of the ensemble, granting it the ability to scale up or down (robustness and scalability). If the swarm is sufficiently able to self-organize, roles will be distributed to serve the needs of the ensemble (flexibility). Lastly, by its nature, such systems can be executed in parallel.

Keeping these properties in mind, we might find that a system of repeated neural modules displays a large amount of plasticity and robustness, and scales up or down depending on the computational needs (could we seamlessly scale up or down the number of sensors in a system? Could increasing size increase compute power?). Moreover, such a system would display an attractive simplicity in terms of optimization (due to few parameters) and function. However, the system would be more difficult to analyze, because the nature of such a system is that its smallest components' function cannot explain the larger, emergent behavior.

Swarm research could inform us on how to integrate information from an ensemble. In their \citeyear{hawkins2019framework} paper, \citeauthor{hawkins2019framework} explains that the cortical columns would vote to reach a consensus on what object they are observing. Likewise, swarms can reach consensus through a voting process, even when lacking centralized processing. Hawkins et al. further describe that the information in the neocortex could be integrated in the Thalamus, which opens for lack of consensus. As has been shown, swarms do not require any individual to have the right opinion and might estimate the right answer anyway. For example, before the term Swarm Intelligence was coined, \citeauthor{galton1907vox} (\citeyear{galton1907vox}) wrote about circa 800 people being able to collectively estimate the weight of an ox, despite the fact that the vast majority were not particularly well suited to make the prediction. Crucially, this system was highly robust to noise (what kind of day was each individual having?), as well as to expertise (both butchers and laymen were sampled), and could happen fast, in parallel (discussion was not encouraged). However, \citeauthor{krause2010swarm} (\citeyear{krause2010swarm}) showed that although this is true for independent and unbiased answers, it can be proven false in situations that prevent "useful information extraction", such as a combinatorial problem. They note that sometimes expert opinions are strictly necessary, and here consensus decisions based on local interaction among individuals can be used. Observed insect methods rely on expert opinions to cascade through the group, until a threshold is reached and the swarm makes a decision \citep{sumpter2009quorum}. In these methods, several insects might have opposing opinions. Still, good ones will have better success in recruiting supporters, meaning that a fully decentralized consensus can be reached based on the quality of the opinion. Even so, there is an observed speed vs. accuracy trade-off, and how a swarm weighs speed or accuracy can change depending on the urgency to make a decision \citep{sumpter2009quorum}. 

When assuming that the resulting AI system or robot will be deployed in the real world, and might sustain injuries or need to reconfigure its body to solve a specific task, swarm properties appear useful. However, they do come with a price to pay, sometimes quite literally. In Modular Robotics, where robots consist of repeated mechanical modules, feature creep is a known problem. Feature creep happens when each module is equipped to create a behavior for the ensemble, which means that pricey components will be repeated throughout the whole ensemble, increasing the cost \citep{seo2019modular}. Each module is larger, more costly, and more expensive than it needs to be. However, in natural swarm systems, we can expect the opposite effect because the ensemble has emergent functionality that cannot be found in the individual swarm agent. The importance of considering the ensemble when optimizing a system of neural module repetition thus becomes apparent: It is likely a poor idea to optimize the individual module without scoring it on ensemble performance. 

One might also question whether redundancy and fault tolerance are helpful outside embodied systems. The downside of having a self-organizing behavior, which can sometimes be less exact and more stochastic \citep{Dorigo2007}, might indeed outweigh fault tolerance when fault tolerance is not needed. The outlined benefits might be more beneficial when a system needs to be tolerant to noise, damage, and embodiment, as opposed to an expert system that only receives curated input. 

\section{The methods}
\label{seq:methods}

Neural module repetition can be found across many different fields in AI, in vastly different ways. In this section, an attempt to group these methods is made. Considering that this review aims to enable work on this topic, special consideration will be given to recounting and summarizing the various optimization methods and degrees of homogeneity and regularity. An overview can be found in tables \ref{methodoverview} and \ref{task_n_type}.

\begin{table*}[hbt!]
\begin{threeparttable}
\caption{\textbf{Overview of optimization methods used.} The "Name" column refers to the name the authors associated with the work, which is why not all papers have one.}
\label{methodoverview}
\begin{tabular}{lll}
\toprule
Article & Name & Optimization method \\
\midrule
\cite{moon2001block} & BbNN & Evolutionary Algorithm  \\
\cite{jiang2007block} & BbNN & Evolutionary Algorithm  \\
\cite{srivastava2015highway} & HighwayNet & Backpropagation \\
\cite{szegedy2015going} & GoogLeNet & Backpropagation \\
\cite{he2016deep} & ResNet & Backpropagation \\
\cite{larsson2016fractalnet} & FractalNet & Backpropagation \\
\cite{zoph2016neural} & NAS & Reinforcement Learning + BP\tnote{a} \\
\cite{liu2017hierarchical} & -& Evolutionary Algorithm + BP\tnote{a} \\
\cite{miikkulainen2017evolvingdeepneuralnetworks} & CoDeepNEAT & Coop-EA\tnote{c} + BP\tnote{a} \\
\cite{liang2019evolutionary} & LEAF & Coop-EA\tnote{c} + MOEA\tnote{b} \space + BP\tnote{a} \\
\cite{zhong2020blockqnn} & BlockQNN & Q-Learning + BP\tnote{a} \\ 
\midrule
\cite{murre1992calm} & CALM & Hebbian learning rule \\
\cite{reisinger2004evolving} & Modular NEAT & Cooperative coevolutionary NEAT \\
\cite{doncieux2004evolving} & ModNet & Evolutionary Algorithm \\ 
\cite{mouret_mennag_2008} & MENNAG & Evolutionary Algorithm \\
\cite{tang2021sensory} & AttentionNeuron & Evolutionary Strategy \\
\cite{pedersen2022minimal} & - & Evolutionary Strategy \\
\midrule
\cite{christensen2006evolution} &  -& Evolutionary Algorithm \\
\cite{pathak2019learning} &  -& Proximal Policy Optimization (PPO) \\
\cite{mousavi_multi-agent_2019} & - & Reinforcement Learning \\
\cite{mordvintsev2020growing} & NCA & Backpropagation \\
\cite{randazzo2020self} & - & Backpropagation \\
\cite{huang2020one} & Shared Modular Policies & Reinforcement Learning \\
\cite{variengien2021} & - & Reinforcement Learning \\
\cite{grattarola2021learning} & GNCA & Backpropagation \\
\cite{nadizar2022collective} & Embodied SNCA & Evolutionary Strategy \\
\cite{kvalsund2022centralized} & - & Evolutionary Algorithm \\
\cite{mertan2023modular} & - & Evolutionary Strategy \\
\cite{kvalsund2024sensor}& ANCA & Evolutionary Strategy \\
\cite{ferigo2025totipotent} & - & Evolutionary Strategy \\
\bottomrule
\end{tabular}
\begin{tablenotes}[hang]
\item[]Acronyms:
\item[a]Backpropagation
\item[b]Multi-Objective Evolutionary Algorithm
\item[c]Cooperative Coevolutionary Algorithm
\end{tablenotes}
\end{threeparttable}
\end{table*}

\begin{table*}[hbt!]
\begin{threeparttable}
\caption{\textbf{Overview of tasks solved and what type of network the module was.} The "Name" column refers to the name the authors associated with the work, which is why not all papers have one. "Type" refers to the type of module network. Given a graph neural network, the type is still given as the module type (typically FFNN).}
\label{task_n_type}
\begin{tabular}{llll}
\toprule
Article & Name & Task & Type \\
\midrule
\cite{moon2001block} & BbNN & Robot control & FFNN \tnote{a} \\
\cite{jiang2007block} & BbNN & Signal classification & FFNN \tnote{a} \\
\cite{srivastava2015highway} & HighwayNet & MNIST, CIFAR-100 & ConvNet \tnote{b}\\
\cite{szegedy2015going} & GoogLeNet & ImagNet & ConvNet \tnote{b} \\
\cite{he2016deep} & ResNet & ImageNet & ConvNet \\
\cite{larsson2016fractalnet} & FractalNet & ImageNet & ConvNet \tnote{b}\\
\cite{zoph2016neural} & NAS & CIFAR-10, Penn Treebank & RNN\tnote{e} \\
\cite{liu2017hierarchical} & -& CIFAR-10, ImageNet & ConvNet \tnote{b}\\
\cite{miikkulainen2017evolvingdeepneuralnetworks} & CoDeepNEAT & CIFAR-10, Penn Tree Bank, & ConvNet \tnote{b}, LSTM \tnote{c}\\
& & MSCOCO & \\
\cite{liang2019evolutionary} & LEAF & Wikidetox, image classification & ConvNet \tnote{b}, LSTM\tnote{c}\\
\cite{zhong2020blockqnn} & BlockQNN & CIFAR-10, ImageNet & ConvNet \tnote{b} \\ 
\midrule
\cite{murre1992calm} & CALM & Classification (Any) & Custom \\
\cite{reisinger2004evolving} & Modular NEAT & Custom board game & FFNN \tnote{a} \\
\cite{doncieux2004evolving} & ModNet & Cart pole, lenticular blimp & FFNN \tnote{a} \\
\cite{mouret_mennag_2008} & MENNAG & Cart pole and robotic arm & FFNN \tnote{a} \\
\cite{tang2021sensory} & AttentionNeuron & Robot Walking, CarRacing, & FFNN\tnote{a} , RNN\tnote{e}\\
 &  & Cart-pole, Pong & \\
\cite{pedersen2022minimal} & - & OpenAI Gym suite & FFNN\tnote{a} , RNN\tnote{e}\\
\midrule
\cite{christensen2006evolution} & - & Reconfiguration & FFNN \tnote{a}  \\
\cite{pathak2019learning} &  -& Robot walking/Standing & FFNN \tnote{a} \\
\cite{mousavi_multi-agent_2019} & - & MNIST & FFNN \tnote{a}, LSTM\tnote{c} \\
\cite{mordvintsev2020growing} & NCA & Static pattern imitation & FFNN \tnote{a}\\
\cite{randazzo2020self} & - & MNIST & FFNN \tnote{a}\\
\cite{huang2020one} & Shared Modular & Robot walking & FFNN \tnote{a}\\
& Policies & & \\
\cite{variengien2021} & - & Cart pole & FFNN \tnote{a}\\
\cite{grattarola2021learning} & GNCA & Dynamic and static  & FFNN \tnote{a}\\
& & pattern imitation & \\
\cite{nadizar2022collective} & Embodied SNCA & Robot walking & Spiking FFNN\tnote{a}\\
\cite{kvalsund2022centralized} &  -& Robot walking & CTRNN \tnote{d}\\
\cite{mertan2023modular} & - & Robot walking & FFNN \tnote{a} \\
\cite{kvalsund2024sensor} & ANCA & MNIST, Fashion-MNIST, & FFNN \tnote{a}\\
& & CIFAR-10 & \\
\cite{ferigo2025totipotent} & - & Robot walking & FFNN \tnote{a} \\
\bottomrule
\end{tabular}
\begin{tablenotes}[hang]
\item[]Acronyms:
\item[a]Feed Forward Neural Network (also called ANN)
\item[b]Convolutional Neural Network (also called CNN)
\item[c]Long Short-Term Memory
\item[d]Continuous Time Recurrent Neural Network
\item[e]Recurrent Neural Network
\end{tablenotes}
\end{threeparttable}
\label{table:task_type}
\end{table*}

\subsection{Architectural module repetition}

Architectural module repetition is by far the most normal of the methods. In many of the areas we will touch on in this section, the literature is too vast to make a complete review of here. Furthermore, because of the lack of parameter sharing, this repetition of modules will be less likely to have all the benefits outlined above. Even so, we include purely architectural repetition to showcase some of the techniques associated with module repetition in general. In turn, we hope that this overview provides interested readers with an idea of how such networks can be implemented. 

In image recognition, modular architectures have been used for years. Typically, these networks are assembled with a small number of repeated blocks and trained end-to-end with a gradient-based method so that each module performs different functions depending on its position in the ensemble. Connecting the modules is manually optimized. Important contributions to this field were Highway Net, which added connections that skipped several layers in information highways \citep{srivastava2015highway}; GoogLeNet, used the Inception module consisting of several parallel filtering layers, stacked sparsely \citep{szegedy2015going}; ResNet, that made the residual block, where the output of a block is the processed input plus the non-processed, \citep{he2016deep}; FractalNet, where architectural elements are repeated recursively and have self-similarity throughout the network \citep{larsson2016fractalnet}. Motivated by the ImageNet competition, articles on repeated block-based convolutional neural networks (ConvNets) are seemingly never-ending. The literature is characterized by not only universities contributing, but also large companies such as Google (e.g. \cite{szegedy2015going}) and Microsoft (e.g. \cite{he2016deep}). 

In opposition to hand-made architectures, Neural Network Architecture Search (NAS) seeks to automatize architecture design. The introductory paper \citep{zoph2016neural} used Reinforcement Learning to train an RNN to produce the architecture as a series of tokens. \citeauthor{zoph2016neural} showed that they could create a cell that worked better than the LSTM cell at a transfer learning task. Notably, the assembled networks found by the RNN (called child models) were trained with gradient descent during optimization. Training each child model was time-consuming, and later papers would try to amend this. For example, in the introductory article of the architecture BlockQNN \citep{zhong2020blockqnn}, they first optimize a block architecture, inspired by modular designs such as ResNet. Validating the block is done by stacking it to a small network architecture and training it end-to-end on CIFAR-10. The winning module can easily be stacked in a deeper network for more complex datasets (f.ex. ImageNet), while still letting the initial architecture search be efficient on time.

In the same vein as NAS, \citeauthor{miikkulainen2017evolvingdeepneuralnetworks} (\citeyear{miikkulainen2017evolvingdeepneuralnetworks}) took an evolutionary approach to architecture design in CoDeepNEAT. They used a cooperative coevolutionary approach, where a population of ConvNet modules and a population of blueprints (of the architecture that stacks modules) evolved together. They state this facilitated module reuse, which they saw in their networks. For every evaluation of a network, the full network was trained before assessment, which resulted in optimization taking very long. However, the authors observed a Baldwin effect, which means that the networks learned to learn faster as the optimization went on. CoDeepNEAT is extended in Learning Evolutionary AI Framework (LEAF) \citep{liang2019evolutionary} by adding multi-objective optimization and hyperparameter optimization. In a different direction, \citeauthor{liu2017hierarchical} (\citeyear{liu2017hierarchical}) used a hierarchical, evolutionary approach, where modules are repeated in an optimized architecture using a novel, indirect encoding. The modules do not inherit weights as they are evolved in this pipeline, but are retrained end-to-end every generation. 

Another evolutionary approach can be found in Block-Based Neural Networks (BbNN). BbNNs consist of a limited number of architectural units that are repeated (although not on any large scale). The weights and architecture are trained at the same time through an Evolutionary Algorithm (EA). They show that their network has fewer parameters than a fully connected multilayer perceptron (MLP), and showed that it worked on pattern classification and control of a Khepera robot \citep{moon2001block}. \citeauthor{jiang2007block} (\citeyear{jiang2007block}) also used it for ECG Signal Classification and introduced gradient descent search to occur with a given probability while the EA was running. This reduced the training time significantly. Other applications and slight variations on the BbNN have also been used, such as \cite{jiang2005ecg, tran2012real, san2013evolvable, shadmand2016new}; and much more! The applications of BbNN are numerous and include classification and control, notably a lot within health sector applications. A notable improvement was replacing EAs with particle swarm optimization and hybrid variants. 

\subsection{Parameter sharing modules}

In the early 2000s, inspired by both work on modular neural networks and ensembles, AI researchers used coevolutionary approaches to evolve neural modules and sometimes blueprints. Typically, the coevolutionary pressure was cooperative, with the modules receiving a score for being in a good ensemble, and if also optimized, the ensemble receiving a score for performing well on a task. The modules do not have to be repeated in such an ensemble, but we will only consider the instance where they were: Modular NEAT, an offshoot of the influential algorithm NeuroEvolution of Augmenting Topologies (NEAT) in the algorithm family of Topology and Weight Evolving Artificial Neural Networks (TWEANNs). Modular NEAT adapts the original NEAT algorithm to coevolve modules and blueprints \citep{reisinger2004evolving}. The blueprints which specific modules should slot into which specific part of the blueprint. Crucially, the same module can be repeated in the blueprint. The authors specified that this was to allow for bilateral symmetry and other beneficial evolutionary traits. Additionally, modules can grow their number of inputs and outputs during optimization. Finally, the blueprints are allowed more time to evolve, letting them adjust to changes in the module population. Modular NEAT and CoDeepNEAT are almost identical in method, but Modular NEAT evolves the weights as part of the module and does not do gradient descent.

\citeauthor{doncieux2004evolving} (\citeyear{doncieux2004evolving}) used a similar method in ModNet, but evolved the blueprint and modules together (one chromosome, not coevolutionary). They also allowed human-crafted modules to enter the module population. Despite allowing for neural module repetition, it does not appear that this happened extensively in their experiments. ModNet was later used to control a bird-like animat \citep{mouret2006incremental}. Following the example of Modular NEAT and ModNet, as well as indirect encodings in EAs, MENNAG implements an encoding with a focus on the repetition of modules in a hierarchy \citep{mouret_mennag_2008}. Being an indirect encoding, neural module repetition happens often, without separating blueprints and modules. Repetition can even happen amongst super-modules in the hierarchy.

Plastic methods like Hebbian Learning and other local learning methods can offer a way to have module specialization, adaptability, and homogeneous modules. An early module using Hebbian Learning was based on the Mountcastle cortical column called Categorizing and Learning Module (CALM) \citep{murre1992calm}. CALM is a singular neural module, carefully engineered, that is designed to be combined in hand-designed assemblies in accordance with connectionist principles. In opposition to the connectionists, CALM is constructed to be modular in such a way as to enable attention to relevant data. The connections between modules are learned with a form of unsupervised Hebbian learning, while the internal connections are hand-tuned and identical. The modules in a network do not have to be of equal dimensions, 
but dimensions are frequently reused, and the internal neurons function identically. CALM has been used in many works in the 90s and 2000s. However, most of them do not neatly fit into the homogeneous modular repetition we are looking for. Despite the modules being very self-similar, they wildly vary in size (and the sizes are not repeated). For example, 
\citeauthor{cho1998evolutionary} (\citeyear{cho1998evolutionary}) 
excitingly uses an EA to evolve the architecture, but also evolve the number of nodes in each module. Nonetheless, work on the CALM module is interesting in this review because it is one of the few explicit modelings of the cortical column in the reviewed work. Other interesting developments on the CALM module include enhancing it to learn sequences 
\citep{koutnik2004single}.

Inspired by neocortical rewiring, the novel work of \citeauthor{tang2021sensory} (\citeyear{tang2021sensory}) showcased an architectural feature called AttentionNeuron that could adapt to changing inputs on the fly. Inputs, scrambled or not, were individually fed into identical units that outputted value, key, and query for the self-attention calculation. This calculation constructed a latent vector that was then used in a policy for behavioral tasks. Because both the self-attention calculation and the distributed input layer are permutation-invariant (with query being independent of the input, see \citep{tang2021sensory} for further explanation), the whole system becomes fully adaptable to permuted input. Permutation invariance was also achieved by \citeauthor{pedersen2022minimal} (\citeyear{pedersen2022minimal}) in a similar approach, where the aggregation of the units was done with summation instead of an attention calculation. Furthermore, the repeated units were slightly more complex: where AttentionNeuron had an LSTM unit for Cart-pole, \citeauthor{pedersen2022minimal} tests both sizable Gated Recurrent Units and 4-layered FFNNs. A similar effect was seen in both papers: The systems could function under online input permutations, but performance degraded with increased frequency of these permutations. The modules needed some time to self-correct and assume their new roles.

\subsection{Disconnected and parallel modules with parameter sharing}

Not all instances of repeated neural module repetition are connected networks. By virtue of using an EA as an optimization, for example, the MENNAG paper featured a network that had three fully separate parts, treating their own inputs, with one repeated super-module \citep{mouret_mennag_2008}. Disconnected modules are also a staple in robot control, especially when it comes to modular robots. Here, we will review disconnected (decentralized/distributed) modular systems, with the condition that they have neural repetition and are part of a system that functions as one. Such a system could be either a connected body or an ensemble of networks.

\begin{figure}
    \centering
    \includegraphics[width=\linewidth]{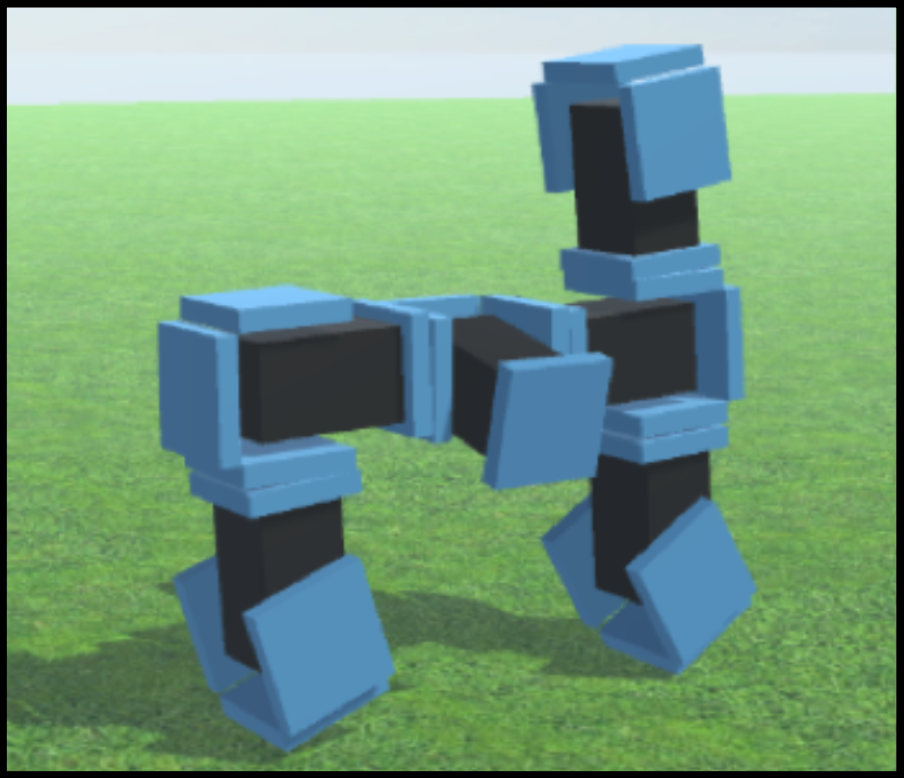}
    \caption{\textbf{An example of a modular robot}. The black and blue physical module is repeated six times to create one creature that can move across a 3D environment. Photo: Part of "Examples of well-performing morphologies" by \cite{kvalsund2022centralized}. Photo licensed under \protect\href{https://creativecommons.org/licenses/by/4.0/legalcode}{CC BY 4.0} } \label{fig:horse}
\end{figure}

Modular robotics concerns robots built from homogeneous, independent mechatronic modules. An example robot can be seen in Figure \ref{fig:horse}. Because modular robots' modules are easily separable and might even function autonomously, the modular robotics field has focused on autonomous reconfiguration as a possibility for these robots. With that, roboticists have long contributed to homogeneously distributed control, and many of them have also interpreted these systems of homogeneous neural networks as swarm agents that form a collective (f.ex. \cite{christensen2006evolution, nadizar2022collective}, and to some extent, \cite{ferigo2025totipotent}). Typically, a network module controls each physical module, and the networks then collaborate on creating collective behavior. Examples of these systems include \cite{christensen2006evolution, pathak2019learning, nadizar2022collective, kvalsund2022centralized, mertan2023modular, ferigo2025totipotent}. Often, these systems are optimized with EAs. However, the notable exception is the article of \citeauthor{pathak2019learning} (\citeyear{pathak2019learning}), who treated each network as its own agent and optimized their shared policy through Reinforcement Learning (and later, \cite{huang2020one} would do the same). In the papers of 
\citeauthor{pathak2019learning} (\citeyear{pathak2019learning}) and \citeauthor{huang2020one} (\citeyear{huang2020one}), networks could pass information to each other. However, they would only receive this information in the next pass through the network. In other words, the networks are disconnected but can give input to each other. 

In a different domain, \citeauthor{mousavi_multi-agent_2019} (\citeyear{mousavi_multi-agent_2019}) used a multi-agent system of up to 6 homogeneous agents to perform image classification on the MNIST data. The agents consisted of image classification, communication modules, and decision and belief LSTM modules. Through traversing the image, observing small parts of it, and communicating with the other agents, each agent formed an opinion of the image that is combined system-wide to produce the classification. Excitingly, this article quite resembles the framework described by \citeauthor{hawkins2019framework} (\citeyear{hawkins2019framework}), since each agent has their own worldview and performs actions to gather more information, as well as being fully homogeneous. With significantly simpler and more numerous agents, \citeauthor{kvalsund2024sensor} (\citeyear{kvalsund2024sensor}) was explicitly inspired by minicolumns and \citeauthor{hawkins2019framework} (\citeyear{hawkins2019framework}). In their system, the Active Neural Cellular Automata (ANCA), each agent contributes to the global classification by traversing the image and continuously outputting beliefs in a message-passing architecture. They showed that the agents displayed attentive behavior by finding and focusing on salient information in the images.   

All networks reviewed in this category share that they are both parallel and pass information to the next iteration. In fact, many of them are special cases of - or share conceptual similarities to - Neural Cellular Automata (NCA) or Graph Neural Cellular Automata (GNCA). NCA concerns the iterated application of one neural network to every neighborhood in a substrate. The substrate is updated iteratively, modeling a dynamically evolving system. As still a very young system, NCAs have so far been used to grow and maintain patterns \citep{mordvintsev2020growing}, classify digits \citep{randazzo2020self}, and even various control tasks \citep{variengien2021}. Extending NCA to graphs \citep{grattarola2021learning}, GNCA are a natural approach whenever the data is graph-structured and dynamically evolving (such as in \cite{pathak2019learning, huang2020one}). Importantly, baseline NCAs and GNCA are end-to-end differentiable, making their training far more energy- and time-efficient than in works based on evolutionary computation. 

\section{Benefits and disadvantages observed in research so far}
\label{seq:observed}

Given the reviewed literature and the theoretical advantages/disadvantages outlined above, this section will focus on the benefits seen by researchers using neural module repetition. The effects must be meaningfully caused by the module repetition. Overall, the selected themes reflect the observed themes in the reviewed works. An exception is made for benchmark scores: Although several papers report a competitive or superior score to state of the art, the goal of this section is to review effects specific to the chosen methods. Of decidedly more interest are the effects that would be useful in any of the major challenges facing AI deployment today, such as the ability to generalize and to compose, its scalability and energy efficiency, as well as its suitability for robotics.  

In addition, although we will not dwell on it, the benefits will differ between the different methods used in the above section. For example, strictly architectural repetition will be absent for the most part, because the reviewed work focused more on benchmark scores. Similarly, disconnected and parallel systems will have a larger degree of swarm-like effects because researchers often created and tested them with swarms in mind. This does not mean that absent works did not display these features, simply that they were not tested for it, since this section is written on claims made by the authors. For full transparency, the citations from the three different methods will be marked, while discussing the themes in general. The marks are as such: Architectural module repetition $\diamondsuit$, parameter sharing modules $\spadesuit$, and disconnected and parallel modules with parameter sharing $\clubsuit$.

\subsection{Generalizable}

Many articles show that their systems are more generalizable than their monolithic counterparts are. \citeauthor{nadizar2022collective} (\citeyear{nadizar2022collective}, $\clubsuit$) showed that their uniform distributed control of robots could adapt better to terrain that had not been seen during training compared to non-uniform distributed control. In their work on Modular NEAT, \citeauthor{reisinger2004evolving} (\citeyear{reisinger2004evolving}, $\spadesuit$) claim that because their neural modules must evolve to be used in so many different parts of the network, their neural modules ends up being more general. This is especially the case because the modules do not necessarily have to communicate with clones of itself, but had to work well with any module in any position to not sabotage the ensemble and hurt its own score. 

This environmental robustness through generalization could be part of the solution to a long-standing problem in co-optimizing robot bodies and brains. In \citeyear{Cheney2016}, \citeauthor{Cheney2016} showed how a robot body and brain, which was co-optimized for the task of walking, had only minimal body development throughout learning. It turns out that because the brain interfaces with the world through the body, changes to the body during optimization was extremely detrimental; it essentially scrambled the controller. However, in the works of  \citeauthor{kvalsund2022centralized} (\citeyear{kvalsund2022centralized}, $\clubsuit$) and \citeauthor{mertan2023modular} (\citeyear{mertan2023modular}, $\clubsuit$), the authors reported that by using repeated modular controllers, the robots were more robust to changes in the body during co-optimization. The ensemble responded well to receiving a new module or losing an old one as part of the optimization, because the controller already adapted to generalize to every position in the body. In other words, it learned to handle many body-parts at once. This meant that body changes were less detrimental compared to centralized and non-modular controllers (\cite{mertan2023modular}, $\clubsuit$), and that such a controller led to more developed and diverse bodies (\cite{kvalsund2022centralized}, $\clubsuit$). In both papers, the repeated modular controllers resulted in an increased task performance, in no small part due to the parameter reduction, but also as a consequence of the increased development of the body.

To further cement that repeated modular controllers are better than monoliths at adapting to changes in the body, \citeauthor{mertan2023modular} (\citeyear{mertan2023modular}, $\clubsuit$) also showed that they are more robust to transfer to unseen bodies. Similarly, \citeauthor{pathak2019learning} (\citeyear{pathak2019learning}, $\clubsuit$) showed that their uniform distributed control could adapt to unforeseen morphologies in zero-shot learning. Specifically, in both walking and standing exercises, the uniform distributed policy could adapt to more modules as opposed to the monolithic policy, which needed to be retrained to handle this case. 

Taking generality to the extreme, \citeauthor{huang2020one} (\citeyear{huang2020one}, $\clubsuit$) shows that a single module policy can control every body part in a large suite of robot bodies in an article aptly titled "One policy to control them all". Continuing the work of \citeauthor{pathak2019learning} (\citeyear{pathak2019learning}, $\clubsuit$), this single policy is trained not only on different perspectives on the task (body parts with varying connections, strengths, and weights), but also on different tasks (different bodies from a distribution). Hence, the policy learns to generalize zero-shot to new bodies, both within and out of distribution (see Figure \ref{fig:concept_figure}g).

Exploring how generality can lead to fault tolerance in the face of body damage, the robot swarm of \citeauthor{christensen2006evolution} (\citeyear{christensen2006evolution}, $\clubsuit$) showed a unique ability to self-repair. His modular robot swarm consisted of up to ca. 3500 modules, and was trained to self-assemble into defined structures. An unforeseen behavior was the swarm's ability to self-repair if modules were removed. He showed that because the modules self-organize with the intent of maintaining its shape, the modules redistribute when modules are taken out in order to still maintain the structure. In short, removing modules will trigger redistribution of modules, which leads to self-repairing the structure. 

In general, the claim of generality is a reoccurring theme in these articles. They see the repeated neural module as a generalist, contrary to a specialist. Although the literature does not necessarily invoke swarms and multi-agent systems, it expresses a similar sentiment of robustness through generality. However, in swarms, the generalist swarm agent usually specializes to take on different roles that create a larger function.

\subsubsection{Specialization}

On top of homogeneous repeated modules, some authors choose to include plasticity. This can lead to specialization, where identical modules take on different roles in the ensemble, leading to diversified behavior. One such work is that of \citeauthor{ferigo2025totipotent} (\citeyear{ferigo2025totipotent}, $\clubsuit$), where Hebbian learning was shown to create different behaviors in modules according to their position in the body. This effect is remarkably similar to the classic role-based controller of \citeauthor{stoy2002using} (\citeyear{stoy2002using}) who displayed that a modular robot with identical modules could be reconfigured, and the controller would work for a new morphology zero-shot. Stoy et al. did not use neural networks; \citeauthor{ferigo2025totipotent} illustrate that similar specialization can be found using only Hebbian learning. 

A similar role-like specialization was observed by \citeauthor{pedersen2022minimal} ((\citeyear{pedersen2022minimal}, $\spadesuit$): In a system that was designed to be permutation invariant, they saw that upon permuting the inputs, modules would switch behaviors completely to maintain system performance, exactly like in \citeauthor{tang2021sensory} (\citeyear{tang2021sensory}, $\spadesuit$). As opposed to \citeauthor{tang2021sensory} (\citeyear{tang2021sensory}, $\spadesuit$), \citeauthor{pedersen2022minimal} ((\citeyear{pedersen2022minimal}, $\spadesuit$) managed to show that the behavioral roles had unique output signatures (see Figure \ref{fig:concept_figure}e), which upon permuting would be fully adopted by another module. Importantly, the system was not trained with permuted inputs and had no plasticity. Role-specialization emerged mainly from the identical modules receiving different input and the integration of the module outputs being permutation invariant. 

\subsection{Scalable}

In the literature, authors sometimes report on the scalability of their systems. For example, the field of Modular Robotics already focuses on scalability because of their goal to increase the sizes of robot ensembles. In one such paper, that of \citeauthor{christensen2006evolution} (\citeyear{christensen2006evolution}, $\clubsuit$), the author experiments with zero-shot tests of scalability, and finds that a system trained on 50 robotic modules scales up to 1200 modules without significant loss in performance. Similarly, \citeauthor{kvalsund2024sensor} (\citeyear{kvalsund2024sensor}, $\clubsuit$) shows that a system trained on 225 modules scales up to 676 modules without any visual change to its behavior. In  both of these works, scalability can lead to resource efficiency, because the system can be trained on a smaller size than it is intended to be deployed at. This is a step in the direction of Green AI \citep{schwartz2020green}, where systems are valued also for their frugality in terms of money, energy, resources, and overall accessibility for smaller research groups.   

It does appear that such scalability still needs research to increase its usefulness. In the case of \citeauthor{christensen2006evolution} (\citeyear{christensen2006evolution}, $\clubsuit$), this scalability means that the modular robot can in theory be trained to create small structures, but be deployed, without loss, to create larger structures -- in practice, there will be loss due to the physical strength of the modules. \citeauthor{kvalsund2024sensor} (\citeyear{kvalsund2024sensor}, $\clubsuit$) remark that although their system and similar architectures are inherently scalable, larger ensembles does not lead to increases in performance. The authors reflect that perhaps the method of integration, which was averaging, suppresses module individuality; and that individuation could allow for more quality-driven collective intelligence consensus between modules. Larger ensembles could then lead to more diverse opinions, which with quality-driven consensus could lead to better performance.  

Especially Neuroevolution (NE) can benefit from the neural repetition's reduction of parameters, making NE a more scalable approach. \citeauthor{reisinger2004evolving} (\citeyear{reisinger2004evolving}, $\spadesuit$) motivates their algorithm, Modular NEAT, by how poorly NE algorithms scale to high-dimensional input. Through neural repetition, they reduce the search space, and therefore NE algorithms can do better. This is generally the case for using indirect encodings, but we can see this effect even in direct encodings featuring repetition, like Modular NEAT and the work of \citeauthor{christensen2006evolution} (\citeyear{christensen2006evolution}, $\clubsuit$), \citeauthor{kvalsund2022centralized} (\citeyear{kvalsund2022centralized}, $\clubsuit$), and 
\citeauthor{nadizar2022collective} (\citeyear{nadizar2022collective}, $\clubsuit$). 

Refreshingly, \citeauthor{moon2001block} (\citeyear{moon2001block}, $\diamondsuit$) go one step further to \textit{prove} that their modular network will always have fewer parameters than a fully connected MLP. Furthermore, they provide a theorem with proof that such a network can always represent an equivalent structure to an MLP, with bounds for number of parameters that allow this. \citeauthor{szegedy2015going} (\citeyear{szegedy2015going}, $\diamondsuit$) similarly reflect that their system, featuring far more repetition than was normal at the time, could achieve training efficiency through sparsity – all while still maintaining the all-too-important depth that the ImageNet contenders were going for. 

\subsection{Architectural constraints}

We identified that repeating modules would necessarily lead to architectural constraints. Throughout the literature, this has been handled in varied ways. We will here point out some common themes in how authors deal with reaching input and output specifications and how authors differentiate module functions.

\subsubsection{Reaching input and output specifications}
Probably the most common method to reach input and output specifications is \textbf{non-modular integration} of module output. Commonly, this can be seen in how the last few layers of ConvNets typically are feed forward layers. Some more novel examples from the reviewed literature can be seen in \citeauthor{tang2021sensory} (\citeyear{tang2021sensory}, $\spadesuit$), where modules are integrated into a latent vector using an attention mechanism. Importantly, the modules themselves help decide the attention, based on the global previous output and current individual input. Another novel example is that of \citeauthor{kvalsund2024sensor} (\citeyear{kvalsund2024sensor}, $\clubsuit$), where an NCA integrates module predictions by essentially averaging them to a global prediction. For a parallel repeated module architecture, integration is a good way to meet output specifications while making the input dimensions dictate the number of modules. This allows the systems to allow a flexible number of inputs without re-training the system, which can be useful for multi-task learning. 

Moreover, we would like to point out that in many of the applications seen, module output is exactly convertible to system output. For example, in the modular robots of \cite{nadizar2022collective, pathak2019learning, kvalsund2022centralized, christensen2006evolution}, $\clubsuit$, and others, module output is directly used for body control because the body and brain are modular one-to-one. Here, repeated modularity is not a constraint, but rather a good way to deal with the repeated modularity constraint of the problem/task. \textit{Because} it allows for flexible numbers of inputs and outputs, repeating module architectures are a natural choice for controlling bodies with changing numbers of modules (\cite{pathak2019learning, kvalsund2022centralized}, $\clubsuit$). A similar effect can be seen in the NCA of \citeauthor{randazzo2020self} (\citeyear{randazzo2020self}, $\clubsuit$), where individual module classifications are directly used to signify system classification.  

\subsubsection{Differentiating module function}
Unanimously, the way to differentiate module function is to supply each module with different coherent subsets of the input. As such, function is differentiated by \textbf{perspectives} on the input. For example, in ConvNets, kernels operate with the same function over the whole input, but each unique pixel neighborhood $n \times m$ generates a unique output $c$. This is the most common denominator in the reviewed work. 

Unique perspectives on the input space also yield different \textbf{episode histories}. Given a system with memory or life-time learning, such as NCAs (\cite{randazzo2020self, variengien2021}, $\clubsuit$), networks with recurrency (\cite{pedersen2022minimal}, $\spadesuit$), or spiking neural networks (\cite{nadizar2022collective}, $\clubsuit$), two modules receiving the same input at time $t$ can output differently depending on their episode histories. As such, module behavior where memory is involved is modulated by life-time specialization. 

Furthermore, \citeauthor{pedersen2022minimal} (\citeyear{pedersen2022minimal}, $\spadesuit$) uses \textbf{random initialization} of hidden states in the modules. If the module is allowed to keep its differentiated hidden state, episode histories might differentiate because of the initialization alone. This is not necessarily ensured, as modules receiving similar input might converge to the same behavior over time, regardless of initialization.

Finally, \citeauthor{randazzo2020self} (\citeyear{randazzo2020self}, $\clubsuit$) employs \textbf{random updates}. Here, an update to the cell memory is only performed with a probability, meaning that even modules with the same perspective and episode history up to time $t$ are unlikely to continue to output the same at time $t+1$. 

An unseen method in the literature to differentiate is that of using variations on the shared parameter set in each module. For example, modules could be sampled from a normal distribution with a shared parameter set $P$ as $P + N(0,\sigma)$. On the same note, Bayesian neural networks, where individual module weights are sampled from parameterized distributions for every forward pass, could be an interesting way to assure module differentiation while keeping parameters the same. Alternatively, modules could be sampled from variations on the parameter set after some space dependent function, $P + F(x, y)$ for module position $(x,y)$. 

\subsection{What about the other benefits?}

In Section 3, on theoretical advantages and disadvantages, many more benefits were listed. However, not everything has been explored in the literature. Especially in the architectural repetition section, the focus of the articles are mainly on competitive scores and the reduction in parameters. The other methods also focus on select attributes, and if a swarm interpretation is used (as in \cite{christensen2006evolution, nadizar2022collective}, $\clubsuit$), the number of effects that can be tested for is limited in any one paper. 

Truthfully, there are a lot of effects that can still be investigated. The literature is very sparse on theoretical and mechanistic explanations for effects of module repetition, and still largely unexplored for empirical results on such effects. For example: 
\begin{itemize}
    \item We see empirical results of generalizing to new perspectives and out-of-distribution problems, but we need better theoretical foundations and mechanistic explanations as to why that is and what its limits may be.  
    \item Although works like \citeauthor{mousavi_multi-agent_2019} (\citeyear{mousavi_multi-agent_2019}, $\clubsuit$) employed integrating outputs in a voting system, classifiers based on swarm voting can still be further explored. We can expect that individual agents in a voting system can be less complex as long as there are many of them, however, this requires emergent modeling.
    \item On the mechanistic side, it could be illuminating to dissect the communication protocols used by message passing systems to solve tasks. What messages are sent? How do they distribute information across the ensemble? Can we understand how consensus is reached through the messages?
    \item Lastly, how can such a system be constructed to enable modules with roles they specialize in? Can functions co-exist within such a system, or will it be affected by the "debugging problem"?
\end{itemize}

\section{The history taken together and the way forward}
\label{seq:conclusion}

This review has explored the history and potential of neural module repetition. Starting in cognitive science and moving into its applications in AI, we explored how modules can be combined/repeated and the current views on their effectiveness. 

The idea of the cortical column as a computational module was established with Mountcastle's work \citep{mountcastle1978organizing, mountcastle1997columnar}. In the 80s, similar ideas on modularity in the brain were perpetuated in cognitive science and AI, when \citeauthor{fodor1983modularity} (\citeyear{fodor1983modularity}) reintroduced the functional and anatomical modularity into cognitive science. Minsky published his book \textit{the Society of Mind}, where the mind was viewed as a collection of modular agents. At the same time, the connectionists and the PDP movement got started \citep{rumelhart1988parallel}, and would adopt the cortical column as the anatomical unit of their distributed systems \citep{murre1992calm}. 

The 80s and 90s were also the start of Convolutional Neural Networks \citep{lecun1989backpropagation}, and Swarm Intelligence got started along with Modular Robots (Then cellular robotics \citep{beni1993swarm, bonabeau1999swarm}). Within EAs in the 2000s, the idea of neural module repetition was being visited in the construction of BbNNs \citep{moon2001block}, networks with limited repetition \citep{reisinger2004evolving, doncieux2004evolving}, and modular robots \citep{christensen2006evolution}.

In the 2010s, deep learning was revived, and the ImageNet competition brought about many interesting examples of neural module repetition. Optimization of architectures was sure to follow in both the birth of the burgeoning field of NAS \citep{zoph2016neural, zhong2020blockqnn} and in more niche solutions like CoDeepNEAT \citep{miikkulainen2017evolvingdeepneuralnetworks}. However, the need for a lot of computing power is notorious in architecture optimization, stemming in large part from the fact that each found architecture is trained with gradient descent during optimization. 

At the end of the past decade, modular robotics started to revisit learned neural module repetition with both RL and EAs. We are seeing the beginning of cataloging the benefits of this approach because it points to more morphological and environmental robustness. As the virtual robotics community struggles with optimizing both the body and the brain at the same time because the brain lacks robustness to changes in the body, a robust and zero-shot adaptable controller could be a way forward. 

In the coming years, there might be more to gather from a different approach to neural networks. Although the current strain of state of the art is showing impressive results, even making headway on AGI benchmarks \citep{chollet2024arc}, we are seeing a worrying trend in terms of resource consumption, scalability, and democratization \citep{bashir2024climate, george2023environmental, thompson2020computational, schwartz2020green, strubell2019energy}. In terms of achieving AI on the level of a living creature, many prominent AI researchers believe that our current methods are not going to be sufficient \citep{risi2021selfassemblingAI, zador2022toward, hiesinger2021self}. \citeauthor{zador2022toward} and \citeauthor{hiesinger2021self} call for a heightened focus on embodiment and seeing intelligence as a result of the situated and historical context it comes from. This will lead us to again try to unite the disparate fields within AI and robot control, as seen in this review, as well as neuroscience and cognitive science. By both borrowing heavily from research on the brain and having been shown to facilitate robustness in performance and bodily optimization, neural module repetition appears to be a useful method going forward. Although the benefits of neural module repetition have only scarcely been looked at, the theoretical advantages of generality, scalability, robustness, and simplicity are promising and warrant more research. At the same time, the methods available to incorporate neural module repetition are, as seen here, rather diverse and numerous.

\bibliographystyle{elsarticle-harv}
\bibliography{bibliography}

\end{document}